\documentclass[]{elsarticle}
\usepackage{epstopdf} %!!!!!!
\usepackage{color}
\usepackage{lscape}
\biboptions{sort&compress}
\usepackage{graphicx}% Include figure files
\RequirePackage{amsmath}
\begin{document}
\begin{frontmatter}
\title{%Investigation of critical behavior of metallic glasses heat capacity near glass-liquid transition: following the studies of shear modulus
Critical behavior of the fluctuation heat capacity  near the glass transition of metallic glasses}

\author[mymainaddress]{R.A. Konchakov}
\author[mymainaddress]{A.S. Makarov}
\author[mymainaddress]{G.V. Afonin}
\author[mymainaddress2]{J.C. Qiao}
\author[mymainaddress3]{M.G. Vasin}
\author[mymainaddress5]{N.P. Kobelev}
\author[mymainaddress]{V.A. Khonik}
\corref{mycorrespondingauthor}
\ead{v.a.khonik@yandex.ru}
\address[mymainaddress]{Department of General Physics, State Pedagogical University, Lenin Street 86, Voronezh 394043, Russia}
\address[mymainaddress2]{School of Mechanics and Civil Architecture, Northwestern Polytechnical University, Xi'an 710072, China}
\address[mymainaddress3]{Vereshchagin Institute of High Pressure Physics, Russian Academy of Sciences,  Moscow 108840, Russia}
\address[mymainaddress5]{Institute of Solid State Physics, Russian Academy of Sciences, Chernogolovka, Moscow district 142432, Russia}

\begin{abstract}
The high-frequency shear modulus of five Zr-, Pd-, Cu-based conventional  and two high-entropy  bulk metallic glasses was measured in a wide temperature range up to the beginning of crystallization. Using these data and general thermodynamic relations, the "fluctuation" heat capacity $\Delta C_f$ determined by local structural fluctuations in the defect regions is introduced and calculated. It is found that $\Delta C_f$ temperature dependence   for all metallic glasses has a large peak located slightly below or above the glass transition temperature but clearly lower than the crystallization onset temperature. The form of this peak resembles the characteristic $\lambda$-peak typical for order-disorder phase transitions. It is suggested that this $\Delta C_f$-peak reflects certain underlying critical phenomenon. The critical temperature $T_0$ (peak temperature) and corresponding critical index $\alpha$ are  determined. Averaged over all seven metallic glasses under investigation in the initial and relaxed states, the critical index  $\textless \alpha\textgreater=0.26$. The results obtained indicate that the fluctuations of thermal energy near the glass transition bear the marks of a continuous phase transition. However, the derived critical index is between those corresponding to a second-order phase transition ($\alpha\approx 0.1$) and  a critical transition characterized by a tricritical point ($\alpha \approx 0.5$).
\end{abstract}

\end{frontmatter}

\section{Introduction}

It is commonly accepted that supercooled liquids at low enough temperatures  inevitably exhibit structural freezing and formation of a solid glassy state, which is known as the glass transition \cite{AngellJChemPhys1972}. Despite decades of intense investigations, the nature of the glass transition and structural relaxation of glasses still remains a major problem of condensed matter  physics \cite{NelsonPhysRevB1983, Nemilov1995, DyreRevModPhys2006, WeiNatCommun2013, AlbertScience2016, SanditovUFN2019, DuMaterToday2020, DongApplPhysLett2021, ShenActaMater2022}.
Currently, quite a few different approaches to a solution of the glass transition problem have been proposed. A wide range of approaches is distributed between the two extreme points of view. On the one hand, it is assumed that the glass transition is a
purely kinetic phenomenon, as evidenced, for instance, by the heating/cooling rate dependence of the glass transition temperature $T_g$. These approaches are summarized in a recent review \cite{SanditovUFN2019}. The second hypothesis is that the glass transition constitutes a specific phase transition or its consequence \cite{NelsonPhysRevB1983,AlbertScience2016}.

The latter point of view is supported by the fact that, despite the enthalpy and specific volume change continuously at $T_g$, their derivatives, the heat capacity and  thermal expansion coefficient, display characteristic jumps at this temperature\footnote{More precisely, the heat capacity displays either a jump or a specific $\lambda$-shape change at $T_g$, depending on whether cooling or heating is employed \cite{DyreRevModPhys2006}, as discussed later.}, which are distinctive of a second-order phase transition \cite{Schmelzer2011}.  Furthermore, various microscopic models and experiments mark the similarity of the glass transition with continuous phase transitions (see Ref.\cite{TournierChemPhysLett2016} and papers cited therein). Perhaps, this is the reason why The International Union of Pure and Applied Chemistry (UPAC) defines the glass transition as a second-order phase transition \cite{IUPAC1997}. However, this definition remains controversial, since glass is not an equilibrium state of matter and, therefore, the glass transition is not a phase transition in the strict sense.

The main difficulty encountered in the theoretical description of the glass transition is the lack of a proper order parameter \cite{ManichevJNonCrystSol1995}.   Among the approaches to solve this problem, one can distinguish the use of the quantitative identification of topological defects in amorphous systems. In this case, the key condition for the phase transition is the existence of nontrivial topologically protected structural excitations (disclinations) in the liquid playing the role of quasiparticles. The glass transition then represents  the condensation in the subsystem of quasiparticles so that the glass transition constitutes a passage between the two states of a system of topological defects, i.e. a state in which these defects are mobile, and a state in which they are frozen \cite{VasinPhysRevE2022,VasinJStatMech2011,VasinPhysicaA2021}. 

The theoretical description of the glass phase as a frozen system of topologically stable defects was actively developed at the end of the last century (see, for example, Refs \cite{NelsonPhysRevB1983,ToulouseCommunPhys1977,RivierPhilMag1979}) and currently again attracts the attention of researchers \cite{BaggioliPhysRevLett2021,BaggioliPhysRevE2022,VasinPhysRevE2022}.
This approach is in line with commonly accepted notions considering defects  as the main elements of glass structure \cite{ChengProgMaterSci2011,HirataScience2013}, which can be described by certain disclination networks \cite{QiPhysRevB1991,BorodinJNocCrystSol1995} characterized by the disclination density tensor \cite{ManichevJNonCrystSol1995}. In addition, it well reproduces the characteristic thermodynamic and kinetic properties of the glass transition such as the Vogel--Fulcher--Tammann law, specific behavior of the susceptibility and non-linear susceptibilities  as well as the appearance of the boson peak in the frequency dependence of the dynamic structure factor near $T_g$ \cite{VasinPhysRevE2022}.

An important features of this topological approach is the ability to describe the singular behavior of the temperature dependence of the heat capacity in the glass transition, which relate it with the continuous phase transitions (see, e.g. Ref.\cite{Nemilov1995}). In the above topological theory, the glass transition is understood as a phase transition in a tricritical point occurring in the disclination system \cite{VasinPhysRevE2022}. According to the theory, the heat capacity temperature dependence upon approaching to the tricritical point $T_0$ from low temperatures (i.e. upon heating) is proportional to the power function \cite{Landau},
\begin{equation}
\Delta C^{heating}_p\propto (T_0-T)^{-\alpha }, \label{DeltaCp}
\end{equation}
where $\alpha $=0.5. This is different from the second order phase transition, for which $\alpha \approx 0.1$. On the other hand, upon crossing of $T_0$ from high temperatures (liquid-to-glass transition), the heat capacity does not have any pronounced singularity but undergoes a jump down,
\begin{equation}
\Delta C^{cooling}_p=C_p^{liquid}-C_p^{glass}=const,
\end{equation}
that agrees with experimental observations of glass transitions \cite{DyreRevModPhys2006,Anderson1984,OjovanAdvCondMatPhys2008}.

In this work, we investigate the behavior of the heat capacity of defect subsystem in the vicinity of the glass transition for several metallic glass-forming alloys, and determine the critical indexes $\alpha $. For this, we analyze the experimental data on the shear modulus of a few metallic glasses (MGs) close to the glass transition in terms of the Interstitialcy theory (IT) \cite{GranatoPRL1992,GranatoEurJPhys2014}, which is extensively applied to the understanding of various relaxation phenomena in MGs (see Refs \cite{KhonikMetals2019,MakarovJETPLett2022} and papers cited therein). The IT assumes that relaxation changes of MGs' physical properties (e.g. shear modulus, enthalpy, entropy and Gibbs free energy) are related to the evolution of glass defect subsystem.

\section{Determination of the heat capacity of the defect subsystem}

\subsection{Interstitialcy theory}

According to the  IT, melting of metallic crystals is associated with rapid generation of interstitial atoms
in the dumbbell configuration, which destabilize the lattice   \cite{GranatoPRL1992,KonchakovJETPLett2021} but remain identifiable structural objects in the liquid state \cite{NordlundEurPhysLett2005} and determine  its thermodynamic properties \cite{GranatoPRL1992,GranatoEurJPhys2014}. Melt quenching causes freezing of the defects, which can no longer be detected in the way used in crystals (two atoms trying to occupy the same potential well) but can nevertheless be identified according to their specific properties (strong sensitivity to the applied shear stress, specific strain fields and low-frequency peculiarities in the vibration spectra, just as in metallic crystals \cite{KhonikMetals2019}).  We call them interstitial-type defects. Heat effects  occurring upon structural relaxation and crystallization of MGs can be interpreted as a result of a change of the defect concentration \cite{KhonikMetals2019}. Since the defects strongly reduce the non-relaxed (high-frequency) shear modulus (diaelastic effect), their concentration can be estimated by means of shear modulus measurements \cite{KhonikMetals2019}. In the course of heat treatment, not only single interstitial-type defects but also their clusters of different sizes coexist in the solid glass \cite{KonchakovJPhysConMatt2019}. Such clusters, including 5-7 interstitial-type defects, form structural elements of a non-crystalline matrix with a dominant icosahedral ordering \cite{KonchakovJPhysConMatt2019}. It is worthy of notice in this connection that an interstitial cluster containing seven dumbbell interstitials in crystals constitutes a perfect icosahedron \cite{IngleJPhysF1981}.

The predictive ability of the IT was confirmed by comparing theoretical calculations of the excess enthalpy and other excess thermodynamic potentials of MGs (i.e. the potentials related exclusively to the solid glassy state) with the results of their determination from calorimetric experiments. Overall, these calculations give all excess thermodynamic potentials equal to their experimental values to a precision of 10-15\%. At that, in the supercooled liquid state (i.e. above $T_g$), the calculated and experimental values of these excess potentials practically coincide \cite{MakarovJETPLett2022}. Therefore, the IT-based concept assuming interstitial-type defects as major elements of a non-crystalline metallic medium clearly indicates that the excess thermodynamic potentials of MGs are determined by the interstitial-type defect subsystem \cite{MakarovJETPLett2022}.

\subsection{Calculation of the heat capacity  using experimental data on shear modulus relaxation}

Thus, there are  solid arguments that the excess thermodynamic potentials of MGs are almost exclusively determined by the subsystem of interstitial-type defects frozen in the solid glass by melt quenching \cite{MakarovJETPLett2022}. This defect subsystem can be characterized by the heat capacity $\Delta C_{def}$. Relaxation changes of the defect subsystem result in a change of the macroscopic glass entropy by $\Delta S_{def}$. This entropy change can be calculated using a general thermodynamic relation \cite{Landau},
\begin{equation}
\Delta S_{def}=-\frac{R_{min}}{T},\label{DeltaSd}
\end{equation}
where $R_{min}$ can be considered as the minimal work, which has to be applied to transform the defect subsystem into a metastable equilibrium  with the glass matrix and $T$ is the absolute temperature. Eq.(\ref{DeltaSd}) shows how much the excess entropy of a non-equilibrium metallic glass differs from its maximal value at a given temperature \cite{Makarov2022JPCM}.

The minimal work $R_{min}$ necessary to rearrange the defect subsystem is related  to the Gibbs free energy barrier  $\Delta \mathcal{G}_{def}$ for elementary transformation in the defect regions \cite{Makarov2022JPCM}, which, in turn, is quantified by the elastic shear resistance of the surrounding medium and that is why conditioned by the instantaneous (high-frequency) shear modulus $G$ (simply shear modulus hereafter). Thus, one can accept that $\Delta \mathcal{G}_{def}=GV_0$, where  $V_0$ is a characteristic volume of local atomic rearrangements in defect regions \cite{Nemilov1968,DyrePhysRevB1996,NemilovJNCS2006,DyreJChemPhys2012}. At that,  relaxation changes of the defect concentration determine the above minimal work and can be related to the \textit{relaxation component} $G_{rel}$ of the shear modulus  $G$  as \cite{Makarov2022JPCM}
\begin{equation}
dR_{min}=\frac{N_A V_0}{\beta}dG_{rel},\label{dRmin}
\end{equation}
where dimensionless $\beta$ (about 20 for different MGs) is the shear susceptibility, which characterizes the dependence of the shear modulus on the defect concentration \cite{KhonikMetals2019} (i.e. defines the diaelastic effect) and $N_A$ is the Avogadro number.

Relaxation rearrangements in the defect subsystem are accompanied by a heat effect, which, using Eqs (\ref{DeltaSd}) and (\ref{dRmin}), can be accepted as \cite{Makarov2022JPCM}
\begin{equation*}
\delta Q=TdS_{def}=-dR_{min}=-\frac{N_AV_0}{\beta}dG_{rel}.
\end{equation*}
Differentiation of this relation over temperature gives the macroscopic heat flow, which very well corresponds to the experimental calorimetric data for different MGs \cite{Makarov2022JPCM}.

The isobaric excess heat capacity related to the defect subsystem then becomes

\begin{equation}
\Delta C_{def}=T\frac{\partial S_{def}}{\partial T}=-\frac{N_AV_0}{\beta}\frac{dG_{rel}}{dT}. \label{Cd}
\end{equation}
As mentioned above, the excess thermodynamic potentials of MGs are conditioned by the interstitial-type defect subsystem. Then, Eq.(\ref{Cd}) shows that the corresponding excess heat capacity $\Delta C_{def}$ of the whole defect subsystem is conditioned by the derivative of the relaxation component of the shear modulus over temperature. Within the IT framework, the heat capacity $\Delta C_{def}$ constitutes the macroscopic excess heat capacity of metallic glass with respect to the maternal crystalline state.

On the other hand, one should recall that elementary rearrangements in defect regions take place as a result of local atomic displacements driven by thermal fluctuations. To characterize these fluctuation-induced rearrangements, one can introduce a "fluctuation" heat capacity $\Delta C_{f}$. According to the fluctuation thermodynamics, the work $R_{min}$ characterizes the change of the Gibbs free energy $\Delta \mathcal{G}_{f}$ determined by fluctuation rearrangements \cite{Landau}.  Then, one can calculate the entropy change due to these rearrangements using a general thermodynamic relation, $\Delta S_{f}=-\left.{\partial \Delta \mathcal{G}_{f}}\right/{\partial T}$ \cite{Landau}. Accepting that $\left.{\partial \Delta \mathcal{G}_{f}}\right/{\partial T}=\left.{\partial R_{min}}\right/{\partial T}$, where $R_{min}$ is given by Eq.(\ref{dRmin}), one arrives at the  heat capacity determined by thermal fluctuations,

\begin{equation}
 \Delta C_{f}=T\frac{d \Delta S_{f}}{dT}=T \frac{N_AV_0}{\beta}\frac{d^2G_{rel}}{dT^2}, \label{Cf}
\end{equation}
It is seen that the fluctuation heat capacity  $ \Delta C_{f}$ is given by the second derivative of the relaxation component of the shear modulus over temperature, contrary to the  defect-induced heat capacity  $\Delta C_{def}$ (\ref{Cd}) determined by the first derivative of the shear modulus relaxation component. It is also to be mentioned  that the defect heat capacity $\Delta C_{def}$ (\ref{Cd}) equals the fluctuation entropy $\Delta S_f$. This is in agreement with general thermodynamic theory of fluctuation, which shows that the mean square fluctuation of the entropy is proportional to the heat capacity of a body \cite{Landau}.

To calculate the relaxation component of the shear modulus $G_{rel}$ entering Eqs (\ref{Cd}) and (\ref{Cf}),  let us represent the shear modulus as a sum of the harmonic, anharmonic, electronic and relaxation components \cite{Makarov2022JPCM,MakarovMetals2022}:
\begin{equation}
G(T)=G_0\;[1-a_{anh}T-a_{el}T^2+a_{rel}(T)]. \label{G}
\end{equation}
where $G_0$ constitutes temperature-independent harmonic contribution to the shear modulus, the term $-G_0a_{anh}T$ gives the anharmonic component, the summand  $-G_0a_{el}T^2$ describes the component of the shear modulus occurring due to the free electrons and $G_0a_{rel}(T)$ corresponds to the relaxation component of the shear modulus, which depends on temperature $T$ and thermal prehistory in a complicated way. Expressing the relaxation part of the shear modulus $G_{rel}$ from Eq.(\ref{G}) and taking into account that the electronic component of shear modulus $-G_0\alpha_{el}T^2$ in Eq.(\ref{G}) is much less than any other component \cite{MakarovMetals2022} and, therefore, can be neglected, one arrives at
\begin{equation}
\frac{dG_{rel}}{dT}\approx \frac{dG}{dT}+G_0 \alpha_{anh}T, \label{dGreldT}
\end{equation}
\begin{equation}
\frac{d^2G_{rel}}{dT^2}\approx \frac{d^2G}{dT^2}. \label{dD2GreldT2}
\end{equation}
These equations show that the derivatives of the relaxation component of the shear modulus
$G_{rel}$ entering relations (\ref{Cd}) and (\ref{Cf}) can be calculated from experimental data on temperature dependencies of the shear modulus $G(T)$. 

\section{Experimental details}

The investigation was performed on  bulk glassy conventional Cu$_{49}$Hf$_{42}$Al$_9$, Zr$_{46}$Cu$_{45}$Al$_7$Ti$_2$, Zr$_{65}$Cu$_{15}$Al$_{10}$Ni$_{10}$, Pd$_{40}$Ni$_{40}$P$_{20}$, Pd$_{43.2}$Cu$_{28}$Ni$_{8.8}$P$_{20}$  and high-entropy Ti$_{16.7}$Zr$_{16.7}$Hf$_{16.7}$Cu$_{16.7}$Ni$_{16.7}$Be$_{16.7}$, Zr$_{35}$Hf$_{17.5}$Ti$_{5.5}$Al$_{12.5}$Co$_{7.5}$Ni$_{12}$Cu$_{10}$ (at.\%)  metallic glasses produced by melt suction in a copper mold. The non-crystallinity of castings was confirmed by X-ray diffraction. Differential scanning calorimetry (DSC) was carried out by a Hitachi DSC 7020 instrument operating in high purity (99.999 \%) nitrogen atmosphere at a heating rate of 3 K/min. The reference cell of the DSC instrument contains crystallized sample of a glass under investigation so that the measured heat flow constitutes the difference between the heat flow coming from glass and its crystalline state.  

The electromagnetic acoustic transformation (EMAT) method \cite{VasilievUFN1983} was used to measure the transverse resonant frequencies $f$ (500-700 kHz) of samples  in a vacuum of $\approx{0.01}$ Pa. For this purpose, frequency scanning was automatically performed every 10-15 s upon heating  and the resonant frequency was determined as the frequency corresponding to the maximal signal response received by a pick-up coil. The shear modulus was then calculated as  $G(T)=G_{rt}f^2(T)/f^2_{rt}$, where $f_{rt}$ and $G_{rt}$ are the  vibration frequency and shear modulus at room temperature, respectively. The errors for the absolute $G_{rt}$-values were accepted to be 1-2$\%$. The errors in the absolute $G(T)$-data are about the same while the error in the measurements of $G(T)$-changes was estimated to be $\approx 5$ ppm near room temperature and about 100 ppm near $T_g$.

\section{Experimental results}

\begin{figure}[t]
\centering
\includegraphics[scale=0.3]{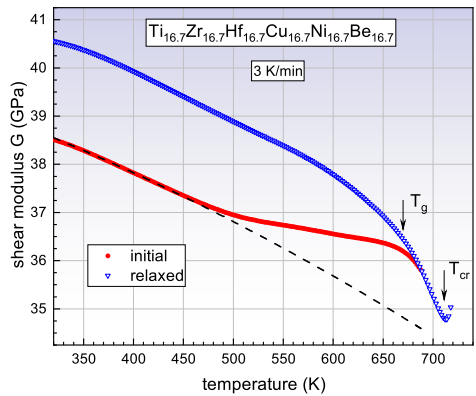}
\caption[*]{\label{Fig1.eps} Temperature dependences of the shear modulus $G$ of bulk glassy high-entropy Ti$_{16.7}$Zr$_{16.7}$Hf$_{16.7}$Cu$_{16.7}$Ni$_{16.7}$Be$_{16.7}$ in the initial state and after relaxation performed by heating into the supercooled liquid state. The calorimetric glass transition temperature $T_g$ and crystallization onset temperature $T_{cr}$ are indicated by the arrows. The dashed line gives linear approximation of $G(T)$-dependence at temperatures $T\leq 450$ K where structural relaxation is absent. }
\label{Fig1}
\end{figure} 

Figure \ref{Fig1.eps} shows temperature dependences of the shear modulus $G$ of high-entropy Ti$_{16.7}$ Zr$_{16.7}$Hf$_{16.7}$Cu$_{16.7}$Ni$_{16.7}$Be$_{16.7}$ glass in the initial state (1st run) and after relaxation (2nd run) performed by heating up to $T=688$ K (deep in the supercooled liquid state) and subsequent cooling to room temperature at the same rate. In the initial state, one observes a linear $G$-decrease at temperatures $T\leq 450$ K due to the anharmonicity approximated by a dashed line, which is followed by an increase of the shear modulus over pure anharmonic decrease due to exothermal structural relaxation. Reaching  the calorimetric $T_g\approx 670$ K leads to a rapid decrease of the shear modulus so that the slope $dG/dT$ increases from $-3.6\times10^{-3}$ GPa/K below $T_g$ up to $-2.6\times10^{-3}$ GPa/K in the supercooled liquid state above $T_g$. Within the IT framework, this shear softening effect and related strong heat absorption can be understood in terms of a rapid multiplication of interstitial-type defects above $T_g$ \cite{KhonikMetals2019}. 

The relaxation by heating up to $T=688$ K  and subsequent cooling results in $\approx 5\%$ increase of the shear modulus at room temperature. Heating of the relaxed sample does not lead to any exothermal structural relaxation and $G(T)$-dependence is nearly linear below 600 K. At higher temperatures, the shear modulus decreases faster than in the case of pure anharmonic behavior, which is due to the endothermal structural relaxation leading to an increase of the defect concentration. Above $T_g$, temperature dependence $G(T)$ in the relaxed state coincide with that in the initial state indicating the loss of the memory of the thermal prehistory in the supercooled liquid region.  

Temperature dependences of the shear modulus for other MGs under investigation are quite similar to that exemplified by Fig.\ref{Fig1.eps} and are not shown here. 

A typical calorimetric behavior of MGs is illustrated by Fig.\ref{Fig2}, which shows a DSC scan of glassy Cu$_{49}$Hf$_{42}$Al$_9$. The supercooled liquid state is manifested by a strong endothermal heat flow, which is followed by sharp crystallization-induced exothermal reaction. It is important to notice that the initial state (1st run) is characterized by a notable exothermal reaction, which disappears after the relaxation (2nd run). Thus, the relaxed sample displays only endothermal relaxation, which is moderate below $T_g$ and quite pronounced above $T_g$.

\section{Calculation of the heat capacities and estimate of the  critical index $\alpha $}

Using the derived $G(T)$-dependences, first, the summands $G_0$ and $G_0\alpha_{anh}T$ in Eq.(\ref{dGreldT}) were determined for all MGs. This was done by linear fitting of these dependences considered in temperature range corresponding to the absence of structural relaxation as exemplified by the dashed line in Fig.\ref{Fig1.eps} (see also Ref.\cite{MakarovMetals2022} for more details). This allowed to derive the relaxation component of the shear modulus as
\begin{equation}
G_{rel}(T)=G_0\alpha_{rel}(T)=G(T)-G_0+G_0\alpha_{anh}T. \label{Grel}
\end{equation}
Then, the first and second derivatives of $G_{rel}$ over temperature were determined (see Eqs ((\ref{dGreldT}) and (\ref{dD2GreldT2})) and the heat capacities $\Delta C_{def}$ (Eq.(\ref{Cd})) and $\Delta C_{f}$ (Eq.(\ref{Cf})) were next calculated. 

\begin{figure}[t]
\centering
\includegraphics[scale=1.8]{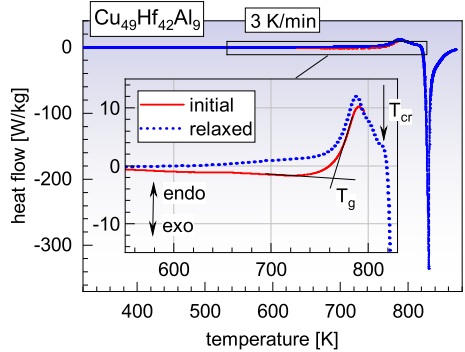}
\caption{DSC trace of glassy Cu$_{49}$Hf$_{42}$Al$_9$ in the initial and relaxed states at a rate of 3 K/min. The inset shows the region below glass transition, the supercooled liquid region  and crystallization onset on an enlarged scale. The ways of the determination of the glass transition and crystallization onset temperatures, $T_g$ and $T_{cr}$, are shown.  }
\label{Fig2}
\end{figure}

\begin{figure}[t]
\centering
\includegraphics[scale=0.55]{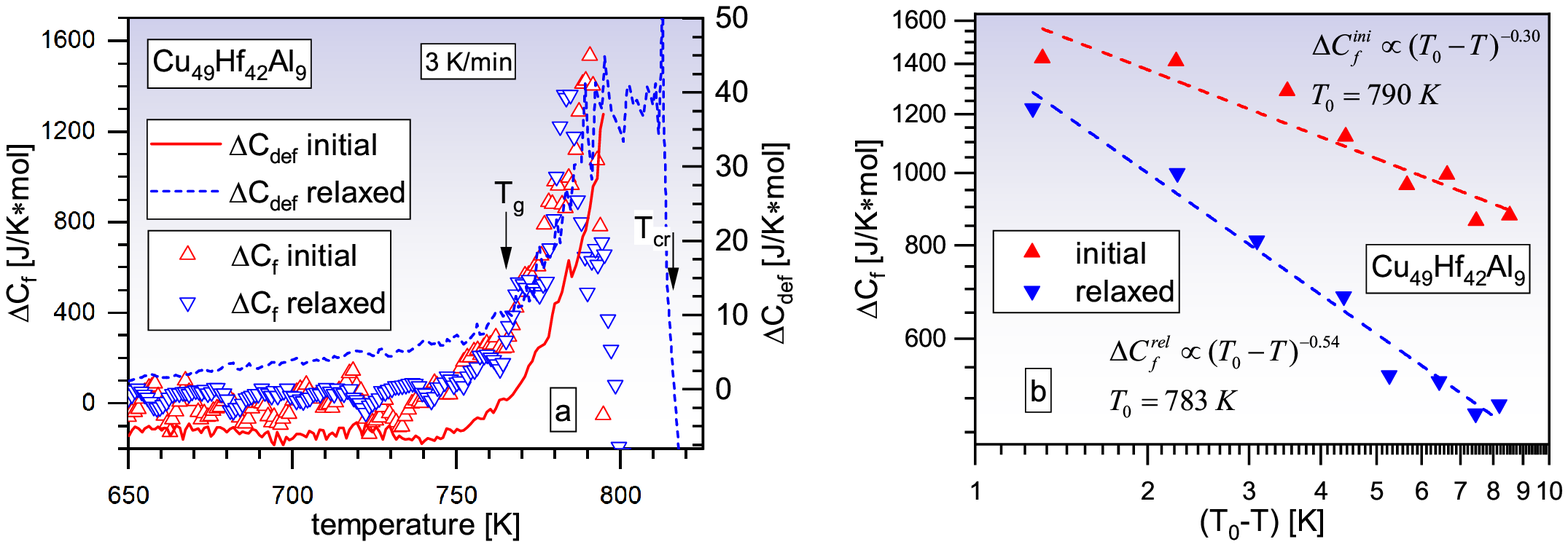}
\caption{Panel (a): Temperature dependencies of the defect heat capacity $\Delta C_{def}$ and fluctuation heat capacity $\Delta C_{f}$ for glassy Cu$_{49}$Hf$_{42}$Al$_9$ in the initial and relaxed state. The glass transition and crystallization onset temperatures, $T_g$ and $T_{cr}$, are indicated. The peaks on $\Delta C_{f}(T)$-dependences are seen. Panel (b): fluctuation heat capacity $\Delta C_{f}$ of the same glass as a function of $T_0-T$ ($T_0$ is the temperature of the $\Delta C_f$-peak) in the logarithmic scale on both axes. The dashed lines give the least-square fits. The critical indexes $\alpha$ and critical (peak)  temperatures $T_0$ are indicated.  }
\label{Fig3}
\end{figure}

Figure \ref{Fig3}(a) illustrates the obtained results using the data on  Cu$_{49}$Hf$_{42}$Al$_9$ glass taken as an example. In the initial state, the defect heat capacity $\Delta C_{def}$ is first negative then starts to rapidly increase just below $T_g$ up to a level of $\approx 40$ J/(K$\cdot$mol). Further heating results in a drop of $\Delta C_{def}$ down to zero above the crystallization onset temperatures $T_{cr}$. The negative $\Delta C_{def}$-values originate from exothermal structural relaxation (see Fig.\ref{Fig2}), which is determined by a decrease of the defect concentration in the initial glass \cite{KhonikMetals2019}. After the relaxation, this effect disappears and $\Delta C_{def}$ is always positive that can be attributed to an increase of the defect concentration and related heat absorption (which is seen in Fig.\ref{Fig2} \cite{KhonikMetals2019}.  Above $T_g$, the defect heat capacity in the initial state is close to that in the relaxed state.  

\begin{figure}[p]
\centering
\includegraphics[scale=0.55]{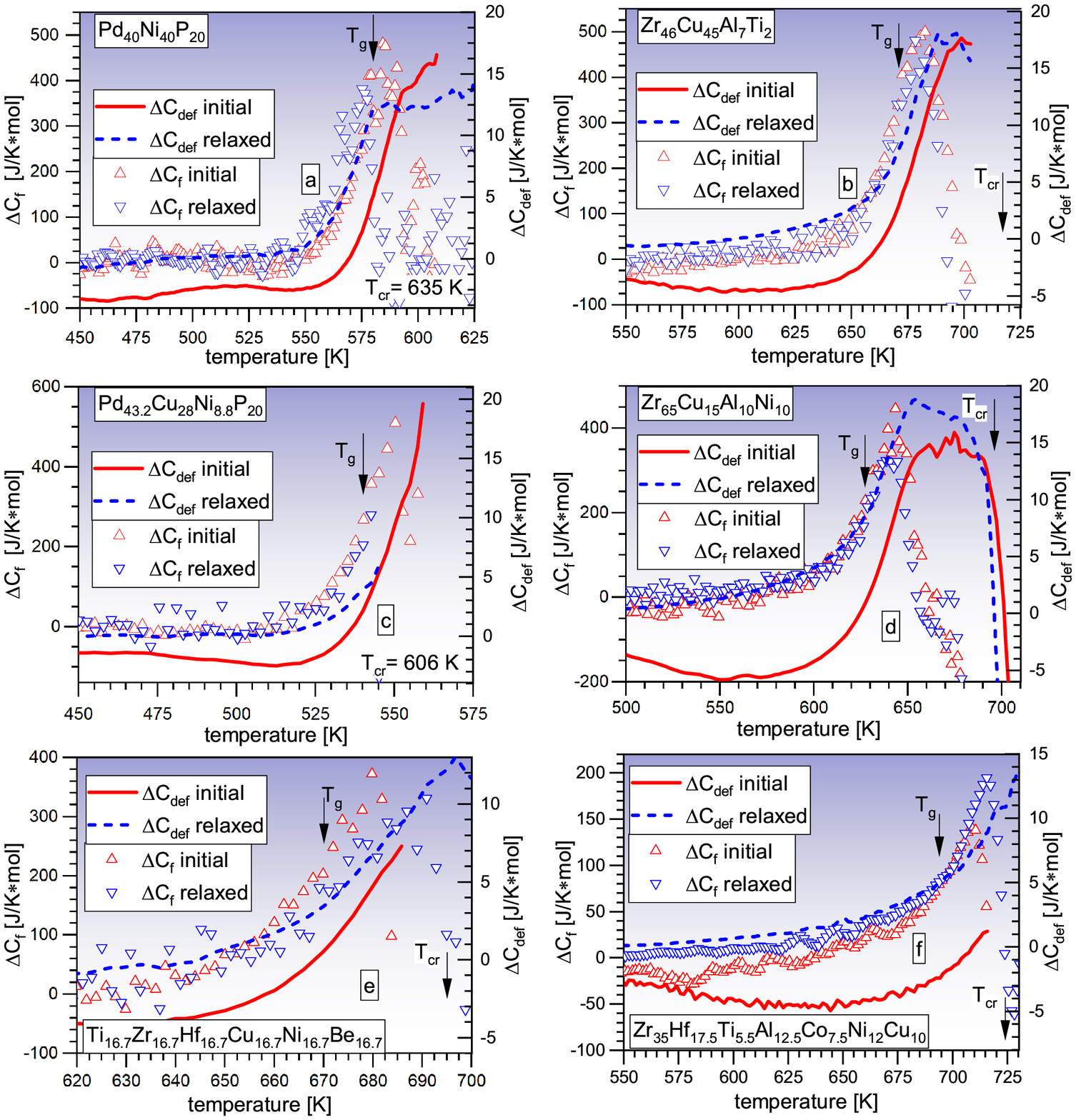}
\caption{Temperature dependencies of the defect heat capacity $\Delta C_{def}$ and fluctuation heat capacity $\Delta C_{f}$ for the indicated  conventional and high-entropy MGs in the initial and relaxed states. The peaks of the fluctuation heat capacity $\Delta C_f$ slightly below or above the glass transition temperature $T_g$ are seen. The $\Delta C_{f}(T)$-peaks are located below the crystallization onset temperatures $T_{cr}$.}
\label{Fig4}
\end{figure}

The fluctuation heat capacity $\Delta C_{f}$ in both initial and relaxed states is first close to zero up to temperatures $T\approx T_g-20$ K, then rapidly increases and  forms a large peak, which is located by about 50 K below the crystallization onset temperature $T_{cr}$. 

Figure \ref{Fig4} shows temperature dependences of the heat capacities  $\Delta C_{def}$ and $\Delta C_{f}$ for all other MGs under investigation. In general, these dependences are quite similar to that shown in Fig.\ref{Fig3}(a). The following common features can be summarized. 

First, the defect heat capacities $\Delta C_{def}$ in the initial state are negative up to temperatures by about 3-15 K below $T_g$ but rapidly increase at higher temperatures. Within the IT framework, these features can be attributed to exo- (endo-) thermal structural relaxation conditioned by a decrease (increase) of the defect concentration and related shear hardening (softening) \cite{KhonikMetals2019}.
 It is to be noted that the maximal $\Delta C_{def}$-values change from $\approx 15$ J/(K$\cdot$mol)   to  $\approx 40$ J/(K$\cdot$mol) (i.e. about 0.6 $R$ to 1.6 $R$, where $R$ is the universal gas constant). As a matter of fact, these maximal $\Delta C_{def}$-values correspond to the increase of the heat capacity upon heating from the glassy to supercooled liquid state (i.e. represent the $\Delta C_{glass-liquid}$ heat capacity increase) and in general agree with literature data. For instance, the maximal defect heat capacity for glassy Pd$_{43.2}$Cu$_{28}$Ni$_{8.8}$P$_{20}$ is about $\Delta C_{def}=20$ J/(K$\cdot$mol) (Fig.\ref{Fig4}(b)) that practically equals $\Delta C_{glass-liquid}\approx 21$ J/(K$\cdot$mol) for a similar Pd$_{42.5}$Cu$_{27}$Ni$_{9.5}$P$_{21}$ glass \cite{NeuberActamater2021}. At that, Zr-based  glasses (Fig.\ref{Fig4}(b,d,f)) show the maximal values $\Delta C_{def}\approx 20$ J/(K$\cdot$mol) while  about the same values of $\Delta C_{glass-liquid}$ heat capacity jump are reported for other Zr-based MGs \cite{JiangActaMater2008}.  
 
The fluctuation heat capacity $\Delta C_f$ in all cases demonstrates a peak at a temperature $T_0$. Table 1 gives the characteristic temperatures $T_g$, $T_{cr}$ and the peak temperatures $T_0$ in the initial and relaxed states. It is seen the $\Delta C_f$-peak is located by 10 K to 63 K below the crystallization onset temperature. It can be concluded, therefore, that this peak is not related to crystallization. It is worthy of notice that the height of this peak is by more that an order of magnitude bigger that the height of the defect-induced excess heat capacity peak.

\section{Discussion}

Temperature dependence of the fluctuation heat capacity $\Delta C_{f}(T)$ derived above reflects the changes of the fluctuation energy  with increasing temperature. Figs \ref{Fig3}(a) and Fig.\ref{Fig4} show that this dependence displays a peak located quite near the glass transition temperature but below the crystallization onset temperature. The form of the peak resembles the characteristic $\lambda$-peak typical for order-disorder phase transitions \cite{Frenkel1946}. The phenomena occurring in the vicinity of phase transitions due the fluctuations of the thermodynamic parameters are generally called the critical phenomena while the transition temperature $T_0$ is referred to as the critical temperature \cite{Landau,Frenkel1946}. The approach to a critical point results in drastic increase of the fluctuations that leads to a (theoretically unlimited) growth of the second derivatives of the Gibbs free energy including the heat capacity. The thermodynamic parameters at critical points have certain peculiarities related to the specificity of fluctuation phenomena \cite{StishovJETP2020}. In particular, the critical points for  second-order phase transitions reflect the abnormal growth of the fluctuations of the order parameter due to the flatness of the Gibbs free energy near the transition point \cite{Landau}. Due to the absence of a discontinuity of the derivative of the Gibbs free energy, second-order phase transitions are also called "continuous phase transitions". The phenomena associated with continuous phase transitions are also called the critical phenomena, due to their association with critical points. 

A curve of a second-order phase transition  separate the phases (on a $(P,T)$ diagram) with different symmetry and cannot terminate in a certain point. This curve, however, can transform into a curve of a first-order phase transition. The corresponding transition point is called the tricritical point \cite{Landau}. Temperature dependence of the change of the isobaric heat capacity near  the critical points is generally described by Eq.(\ref{DeltaCp}). The critical index $\alpha$ in this relation reflects the origin of the phase transition. For a second-order phase transition, $\alpha\approx 0.1$ \cite{GuillouPRB1980,MartynovTVT2018} while for a critical phenomenon corresponding to a tricritical point $\alpha\approx 0.5$ \cite{VasinPhysRevE2022,Landau,StishovJETP2020}. In the latter case, fluctuations are less significant but nonetheless play an important role \cite{StishovJETP2020}.  

\begin{table}[t]
\footnotesize
\caption{Glass transition temperatures $T_g$, crystallization onset temperatures  $T_{cr}$ and peak (critical) temperatures $T_0$ in $\Delta C_f(T)$-dependences (see Fig.\ref{Fig3}(a) and Fig.\ref{Fig4}) together with the critical exponents $\alpha$ for the MGs under investigation in the initial and relaxed states. } 
\begin{tabular}{clcccccc}
\hline
\hline
N & Glass composition (at.\%)& $T_g$ & $T_{cr}$ & $T_0^{ini}$ & $\alpha^{ini}$ & $T_0^{rel}$ & $\alpha^{rel}$\\
  & 	  & [K]   & [K]   & [K]         &                & [K]         &\\
\hline
1 & $Cu_{49}Hf_{42}Al_{9}$ & 765 & 816 & 790 & 0.30$\pm$0.05 & 783 & 0.54$\pm$0.04\\
2 & $Zr_{46}Cu_{45}Al_{7}Ti_{2}$ & 671 & 717 & 677 & 0.27$\pm$0.05 & 676 & 0.26$\pm$0.08\\
3 & $Pd_{40}Ni_{40}P_{20}$ & 580  & 635 & 585 & 0.35$\pm$0.05 & 576 & 0.19$\pm$0.03\\
4 & $Pd_{43.2}Cu_{28}Ni_{8.8}P_{20}$ & 540 & 606 & 550 & 0.32$\pm$0.10 & 543 & 0.26$\pm$0.08\\
5 & $Zr_{65}Cu_{15}Al_{10}Ni_{10}$ & 627 & 696 & 643 & 0.18$\pm$0.07 & 640 & 0.26$\pm$0.08\\
6 & $Ti_{16.7}Zr_{16.7}Hf_{16.7}Cu_{16.7}Ni_{16.7}Be_{16.7}$ & 670 & 695 & 681 & 0.18$\pm$0.04 & 678 & 0.21$\pm$0.05\\
7 & $Zr_{35}Hf_{17.5}Ti_{5.5}Al_{12.5}Co_{7.5}Ni_{12}Cu_{10}$ & 694 & 724 & 707 & 0.17$\pm$0.04 & 714 & 0.16$\pm$0.04\\
\hline
\hline
\end{tabular}
\end{table}

Thus, the  critical index  $\alpha$ provides certain information on the nature of corresponding critical phenomenon. In our investigation, we analyze the heat capacity $\Delta C_f$ conditioned by thermal fluctuations. To obtain  information on related processes, on can plot the function $\Delta C_f(T)$ in the logarithmic scale on both axes. Provided that this data representation is linear, one can determine the critical index $\alpha$. An example for Cu$_{49}$Hf$_{42}$Al$_9$ glass is given in Fig.\ref{Fig3}(b), which shows $ln\;\Delta C_f$ as a function of $ln\;(T_0-T)$ in a small temperature range of $T_0-T\leq 10$ K below $T_0$, where $T_0$ corresponds to the $\Delta C_f(T)$-peak in Fig.\ref{Fig3}(a) and accepted to be the critical point. It is seen that this representation is indeed linear providing a critical index $\alpha\approx 0.30$ in the initial state. Quite surprisingly, the critical index is almost twice larger for the relaxed state, $\alpha \approx 0.54$. This clearly indicates that the fine structure of glass in the supercooled liquid state can be dependent on the thermal prehistory despite the facts that \textit{i}) $\Delta C_f(T)$-dependences for the initial and relaxed states (Fig.\ref{Fig3}(a)) look quite similar (although quite scattered) and \textit{ii})   the shear modulus behavior in the supercooled liquid region  for the initial and relaxed states is also fairly similar (Fig.\ref{Fig1}, just the same was documented earlier for another metallic glass \cite{MakarovMetals2022}). 

Similar results were obtained for other MGs. In all cases, the dependences $\Delta C_f$ vs $T_0-T$ in the log-log scale were found to be linear and the corresponding critical indexes $\alpha$ were calculated. The results are summarized in Table 1, which lists $\alpha$-values for the initial and relaxed states. On the average over all MGs, $\textless \alpha^{ini}\textgreater=0.25$ for the initial state that within the error coincides with $\textless \alpha^{rel}\textgreater=0.27$ for the relaxed state. Most often, $\alpha$-values for the initial and relaxed stated are nearly equal while some exceptions exist (see lines 1 and 3). Ignoring the distinction between different MGs and their structural states, one can accept that the critical index $\alpha\approx 0.26$. This is approximately between a second-order transition ($\alpha\approx 0.1$) and critical behavior near a tricritical point ($\alpha\approx 0.5$). It is also worthy of notice that the critical indexes for high-entropy MGs (lines 6 and 7 in Table 1) tend to the values characteristic of a second-order phase transition that could reflect the high-entropy nature of these glasses.        

The obtained results demonstrate the singular behavior of the fluctuation heat capacity near the glass-to-liquid  transition and confirm the assumption that the glass transition is a continuous phase transition. In this case, the term "phase" is used in a broader sense than usually by considering the glassy state of matter as a phase. At the same time, these result are unexpected. On the one hand, the derived critical index significantly bigger than that for a a second order phase transition.  On the other hand, it is notably less than the $\alpha$-value characteristic of a transition with a tricritical point as predicted in Ref.\cite{VasinPhysRevE2022}.

To explain the obtained the result, one should note that in the case of the tricritical behavior the equality $\alpha =0.5$ is true only in the mean field approximation. In real systems, the accounting for the fluctuation corrections at the tricritical point can lead to $\alpha < 0.5$ \cite{Vasiliev2004}. Besides that, it is known that the calculated critical index is true only in the asymptotic approximation. In a real situation, however, the critical indexes depend largely on the details of the model. In addition, one should not forget that the glassy state is not an equilibrium one and the relaxation dynamics may lead to additional corrections to the critical indexes. Therefore, the question about the universality class of the non-crystalline systems under study remains open for further experimental and theoretical studies.

In any case, one can expect that a "phase" transition might exist provided that the defect subsystem in a metallic glass can be characterized by certain ordering, e.g. a correlation between orientations of the defects. This seems reasonable when taking into account that interstitial-type defects under consideration have certain orientations and interact through their stress fields. This ordering weakens upon approaching to the critical temperature $T_0$. One can roughly estimate the correlation radius $R_c$ of the interaction between the defects  using the relation \cite{YereminPolym1991}

\begin{equation}
  \frac{\left(2R_c\right)^3}{V_0}=\frac{\Delta C_f(T_0)}{\Delta C_{def}(T_0)},
    \end{equation}  
where $V_0$ is the characteristic volume (already mentioned above) comprising atomic rearrangements in defect regions. Direct  comparison of the shear viscosity and shear modulus for a metallic glass above $T_g$ showed that $V_0$ is about one  volume per atom $V_{at}$ \footnote{This estimate was accepted as the average $V_0$-value calculated using shear modulus and viscosity data in the supercooled liquid state  for five MGs  given in Ref.\cite{MakarovJPCM2021}.}. For the data shown in Fig.\ref{Fig4}, the ratio $\frac{\Delta C_f(T_0)}{\Delta C_{def}(T_0)}$ is about 15 to 30 .  Thus, the mutual coordination (ordering) of defects should take place in the volume of about  $(15\div 30)\;V_{at}$. It is to be noted also that the $\Delta C_{f}$-peak for $Cu_{49}Hf_{42}Al_{9}$ glass (Fig.\ref{Fig3}(a)) is several times larger that those for other MGs (Fig.\ref{Fig4}) while the ratio $\frac{\Delta C_f(T_0)}{\Delta C_{def}(T_0)}\approx 40$ is also notably larger.  Thus, this glass is peculiar in some sense that could reflect larger defect ordering, which, in turn, requires larger energy for its destruction. It is interesting to underline in this context that the critical temperature $T_0$ for this glass is about 790 K (Fig.\ref{Fig3}) that practically equals to the temperature of the heat flow peak in the DSC diagram given in Fig.\ref{Fig2}.

\section{Conclusions}

We performed detailed measurements of the high-frequency shear modulus $G$ of seven bulk metallic glasses, including two high-entropy MGs. On this basis, using the general approach given by the Interstitialcy theory, we calculated the heat capacity $\Delta C_{def}$ of interstitial-type defect subsystem of glass. This heat capacity practically equals to the excess heat capacity of glass with respect to the maternal crystalline state. 

Next, we introduced and calculated the "fluctuation"  heat capacity $\Delta C_f$ determined by local structural fluctuations in defect regions. It is shown that while the defect heat capacity  $\Delta C_{def}$ is proportional to the first derivative of the relaxation component of the shear modulus $G_{rel}$ over temperature, the fluctuation heat capacity is determined by the second derivative of $G_{rel}$.   

It is found that the fluctuation heat capacity has a peak near the glass transition temperature but clearly below the crystallization onset temperature. The height of $\Delta C_f$ is 15-40 times larger that the level of the defect heat capacity in the supercooled liquid state while its form resembles 
the characteristic $\lambda$-peak typical for order-disorder phase transitions. It is argued that this $\Delta C_f$-peak corresponds to certain underlying critical phenomenon and the peak temperature $T_0$ represents a critical temperature.  

For all MGs under investigation in the initial and relaxed states, the averaged  critical index is found to be  $\textless \alpha\textgreater=0.26$. The results obtained indicate that the fluctuations of thermal energy near the glass transition bear the marks of a continuous phase transition. However, the critical index is between those corresponding to a second-order phase transition ($\alpha\approx 0.1$) and  a critical transition characterized by a tricritical point ($\alpha \approx 0.5$). It is suggested  that the origin of the observed critical behavior is related to the ordering of the defect structure below $T_0$ and the disruption of this order upon approaching $T_0$. Any similar studies on MGs are unknown to us.

  \section{Acknowledgments}
The work was supported by  Russian Science Foundation under the project No. 23-12-00162.

\end{document}